\documentclass[journal=ancac3,manuscript=article]{achemso}

\usepackage{amsmath,amssymb,amsfonts}
\usepackage{bm}
\usepackage{color}
\usepackage{graphicx}
\usepackage{float}
\usepackage{wrapfig}
\usepackage{verbatim}
\usepackage[english]{babel}
\usepackage[colorlinks=true,bookmarks=false,citecolor=blue,urlcolor=blue]{hyperref} 
\usepackage{natbib}
\usepackage{subfigure}
\usepackage{amstext, mathrsfs, textcomp}
\usepackage{multirow}
\usepackage{xcolor}
\usepackage{enumerate}
\usepackage{epstopdf}
\usepackage{tabularx}
\makeatletter

\usepackage{gensymb}

\setkeys{acs}{etalmode=truncate,maxauthors=0}

\title{Reconfigurable topological states \\ in arrays of bianisotropic particles}

\author{Zuxian He}
\affiliation{State Key Laboratory of Integrated Optoelectronics, College of Electronic Science and Engineering, International Center of Future Science, Jilin University, 2699 Qianjin Street, Changchun 130012, China} 
\altaffiliation{Authors who contributed equally to this work}

\author{Daniel A. Bobylev}
\affiliation{School of Physics and Engineering, ITMO University, Saint Petersburg, Russia}
\altaffiliation{Authors who contributed equally to this work}

\author{Daria A. Smirnova}
\affiliation{Research School of Physics, Australian National University, Canberra ACT 2601, Australia}
\alsoaffiliation{Institute of Applied Physics, Russian Academy of Science, Nizhny Novgorod, Russia}
\altaffiliation{Authors who contributed equally to this work}

\author{Dmitry V. Zhirihin}
\affiliation{School of Physics and Engineering, ITMO University, Saint Petersburg, Russia}

\author{Maxim A. Gorlach}
\affiliation{School of Physics and Engineering, ITMO University, Saint Petersburg, Russia}
\email{m.gorlach@metalab.ifmo.ru}

\author{Vladimir R. Tuz}
\affiliation{State Key Laboratory of Integrated Optoelectronics, College of Electronic Science and Engineering, International Center of Future Science, Jilin University, 2699 Qianjin Street, Changchun 130012, China}  


\begin{document}

\maketitle

\begin{abstract}
Topological photonics promotes a novel approach to resilient light manipulation by exploiting spatio-temporal symmetries of the system and dual symmetry of  electromagnetic field. Various prospective device applications pose the need to flexibly control robust field localization associated with topological modes. Here we design a topological array of resonant dielectric meta-atoms with bianisotropic response induced by a spatial symmetry reduction. Mutual orientation of the designed meta-atoms encodes a staggered bianisotropy pattern capable of trapping topological states in a one-dimensional array containing a small number of particles. We show that the topological interface state can be tailored by the  rotation of coaxial bianisotropic particles arranged in an equidistant lattice. Our experimental implementation based on ceramic horseshoe-shaped disks demonstrates remarkable reconfigurability of electromagnetic topological states.
\end{abstract}

\hfill \break

Photonic topological states uncover a vast range of functionalities enabling resilient routing and localization of light.\cite{Lu2014,Lu2016,Khanikaev17,Ozawa_RMP,Xie2021} A remarkable robustness of these modes to disorder is rooted to the fact that their existence is protected by the global symmetries of the structure which are insensitive to local defects. During the recent years, the concepts of topological photonics have been extensively developed theoretically and implemented experimentally across the entire electromagnetic spectrum from microwaves to the visible.\cite{Wang-Soljacic,Hafezi:2013NatPhot,Rechtsman,Barik,Yang-Chen,Sievenpiper} In all these realizations, lattice symmetries provide the crucial ingredient underlying the topological protection.

Since lattice geometry is difficult to reconfigure, the dynamic control and tuning of topological states localization are challenging. Recently, there were several theoretical proposals to tune the topological states using liquid crystal background medium,\cite{Shalaev2018,Hu2021} integration with transparent conducting oxides,\cite{Kudyshev2019oxides} or exploiting thermo-optical effect in silicon and phase-change materials.\cite{Kudyshev2019termal} However, experimental implementations of tunable topological states are quite limited with few demonstrations available involving nonlinear varactor diodes operating in the radiofrequency range,\cite{Hadad-Nature,Dobrykh} temperature-tunable grating heterostructures\cite{Li2017} or optically controlled topological photonic crystal slabs tuned via pump-induced carrier generation.\cite{Shalaev2019}

In this paper, we point out another route to continuously and precisely tune the topological edge states demonstrating our proposal in proof-of-principle experiments. Specifically, we design a one-dimensional array consisting of the meta-atoms with broken spatial inversion symmetry. Such symmetry reduction gives rise to the bianisotropic response of the meta-atom whereby an incident electric field couples to the magnetic dipole moment and magnetic field generates nonzero electric dipole moment of the meta-atom.\cite{Serdyukov,Alaee-Rockstuhl,Odit_ApplPhysLett_2016,Slob-NP,Asadchy2018, Evlyukhin_PhysRevB_2020} Importantly, interaction of two bianisotropic meta-atoms is sensitive not only to the distance between them but also to their mutual orientation which allows one to tune the topological phase for the fixed lattice geometry. While our initial theoretical proposal\cite{Bobylev2021} assumed topology switching by flipping over the meta-atoms, here we make the next conceptual step by showing the continuous tunability of  topological states manifested through the gradual change in the localization length. 

\section{Design of topological structure}

\begin{figure}[t!]
\centering
\includegraphics[width=1\textwidth]{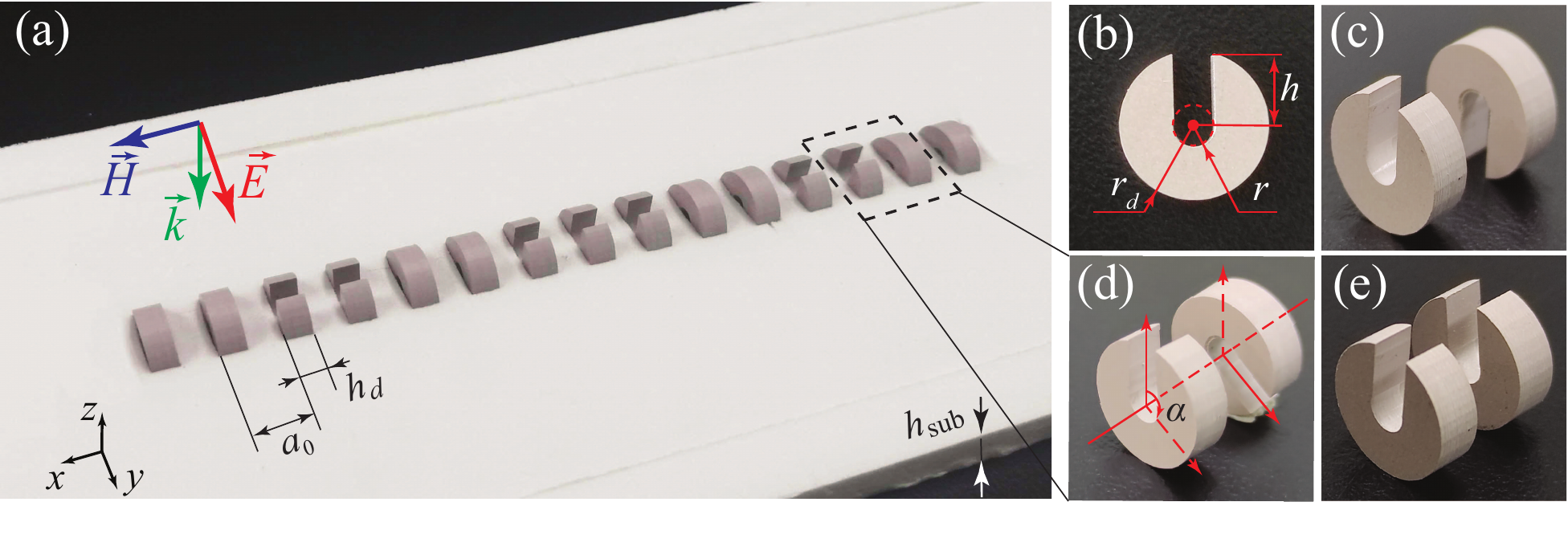}
\caption{{\bf Experimental structure and possible alignments of the meta-atoms.} (a) The photograph of the sample consisting of 15 horseshoe-shaped particles of height $h_d=3$~mm and radius $r_d = 4.5$~mm arranged with the period $a_0=7$~mm. The substrate thickness is $h_\text{sub} = 2.5$~mm. The axis of the array is parallel to the $x$-axis. The particles are made of low-loss high-index ceramics with permittivity $\varepsilon = 24.8$ and dielectric loss tangent $\tan \delta \approx 10^{-3}$. The material parameters of ceramics are referred to 10 GHz frequency. (b) Meta-atom shape and sizes. The inner cutout radius is $r=1.25$~mm, cutout height $h=4.32$~mm. (c-e) Different mutual orientations of the meta-atoms characterized by the angle $\alpha$ between cutout directions: (c) $\alpha = 180\degree$, (d) $\alpha = 150\degree$, and (e) $\alpha = 0\degree$.}
\label{fig:fig1}
\end{figure}

To demonstrate a continuous tuning of topological states experimentally, we design and fabricate a one-dimensional (1D) equidistant array of non-centrosymmetric particles (meta-atoms) (Figure~\ref{fig:fig1}a). The individual meta-atoms are realized as horseshoe-shaped high-index ceramic resonators (Figure~\ref{fig:fig1}b). The array is partially embedded in an air-like ($\varepsilon_\textrm{sub}=1.09$) machine-processed styrofoam supporting substrate which allows us to flexibly rotate the meta-atoms varying their mutual orientation quantified by the rotation angle $\alpha$  (Figures~\ref{fig:fig1}c-\ref{fig:fig1}e). Our experimental studies are supported by full-wave numerical simulations with the use of commercial COMSOL Multiphysics electromagnetic solver.

\begin{figure}[t!]
\centering
\includegraphics[width=1\linewidth]{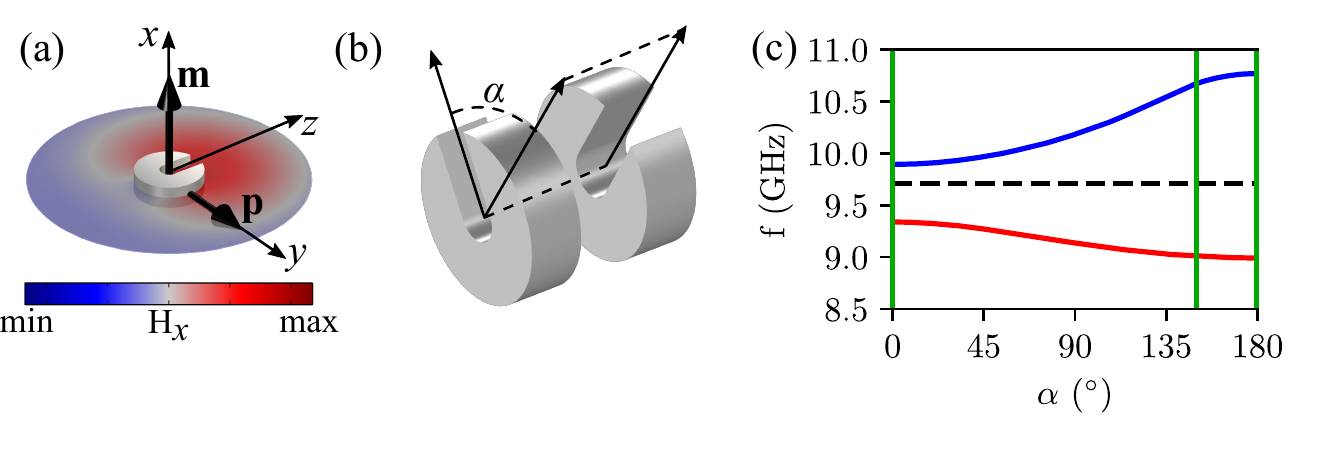}
\caption{{\bf Orientation-dependent coupling of the meta-atoms.}  (a) Hybrid eigenmode of the individual non-centrosymmetric disk disposed in air. The mode features nonzero electric $\mathbf{p}$ and magnetic $\mathbf{m}$ dipole moments simultaneously. (b) Dimer geometry. Angle $\alpha$ allows to tune the magnitude of coupling. (c) Splitting of the meta-atom hybrid magneto-electric mode versus rotation angle $\alpha$ in a dimer. Red and blue curves correspond to the symmetric and antisymmetric modes, respectively. Green vertical lines correspond to the tuning angles ($0^{\circ}$, $150^{\circ}$, and $180^{\circ}$) considered in the experiment.}
\label{fig:fig2}
\end{figure}

To ensure the sufficient tunability of the topological states, we have optimized the design of the individual meta-atom using the multipole expansion technique (see Supporting Information for the details of optimization). For the designed meta-atoms, we choose to work in the vicinity of the lower-frequency hybrid magneto-electric dipole mode with eigenfrequency $f_0 = \left(9.75 - 0.8i\right)$ GHz and field distribution depicted in Figure~\ref{fig:fig2}a.

To quantify the degree of tunability, we simulate the splitting of the resonant frequency into symmetric and antisymmetric modes for a pair of meta-atoms as a function of their relative rotation $\alpha$ for the fixed distance between them (Figure~\ref{fig:fig2}b). Note that the magnitude of splitting measures the effective coupling between the meta-atoms. Therefore, the results of simulations suggest that the magnitude of coupling can change up to 3 times depending on the rotation angle (Figure~\ref{fig:fig2}c). This opens ample options to induce the bandgap, manipulate the topological properties, and change the localization of topological states exploiting orientation degree of freedom. Specifically, the configuration shown in Figure~\ref{fig:fig1}a realizes the scenario with the alternating coupling amplitudes and two repeated weaker coupling links in the middle of the array. As a result, this corresponds to the problem of topological interface states in the well-celebrated Su-Schrieffer-Heeger (SSH) model\cite{Su,Malkova:09,Blanco-PRL} and this mapping is confirmed by our theoretical analysis in the Supporting Information.

\section{Observation of tunable topological states}

The assembled structure was excited by the plane-wave-type field distribution produced by the wideband horn antenna placed in the far-field region. The polarization of the incident field depicted in Figure~\ref{fig:fig1}a was chosen to match the profile of the meta-atom hybrid modes Figure~\ref{fig:fig2}a. The associated near-field profiles measured in the frequency range of $9-10$ GHz were acquired by the subwavelength electric probe that moved across the entire structure.

\begin{figure*}[t!]
\centering
\includegraphics[width=1\textwidth]{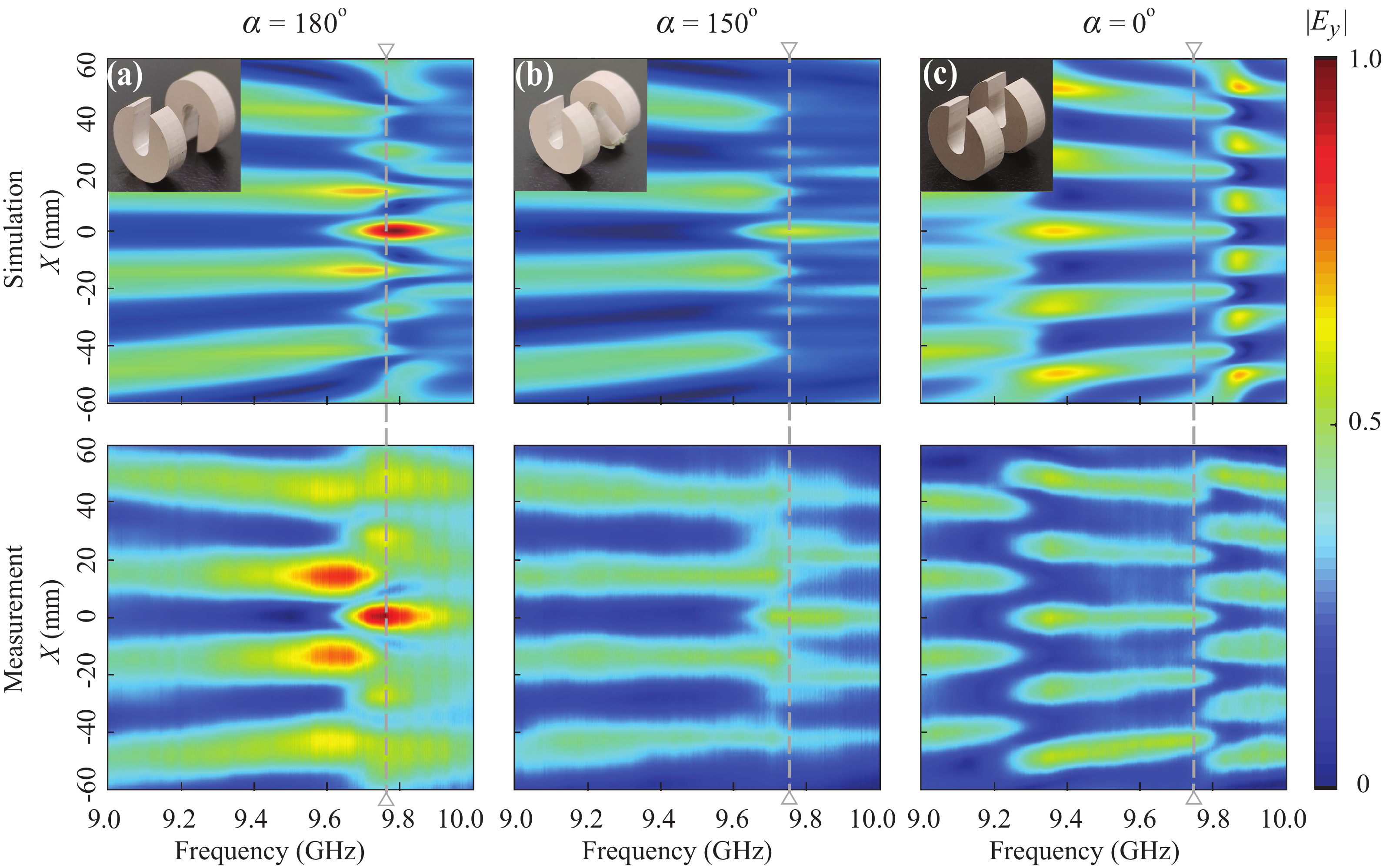}
\caption{{\bf Simulated and measured electric near-field patterns in the frequency range of $9-10$~GHz.} The results are retrieved for the three array configurations with the different orientations $\alpha$ of the meta-atoms. The magnitude of the $y$-component of the electric near-field $|E_y|$ is simulated (top row) and measured (bottom row) at the distance of $1.5$~mm above the structure plane. The frequency of interest ($9.75$ GHz) is highlighted with the gray dashed lines and arrows. Parameters of the array are indicated in the caption to Figure~\ref{fig:fig1}.}
\label{fig:fig3}
\end{figure*}

The results of the electric near-field measurements over the whole array are shown in Figure~\ref{fig:fig3}. Firstly, we investigated the case $\alpha=180^\circ$ with the oppositely oriented particles (Figure~\ref{fig:fig1}a). Both simulated and measured electric near-field patterns (Figure~\ref{fig:fig3}a) exhibit a sharp change in the field localization happening in the frequency interval between $9.7$ and $9.8$~GHz. The localized field pattern indicates the formation of a topological interface state allowing to estimate its localization length.

Then we adjust the geometry of the array by rotating several meta-atoms by $\alpha=150^\circ$ instead of $180^\circ$. Figure~\ref{fig:fig2}c suggests that such a rotation affects the ratio of the two different coupling strengths in the array thus changing the width of the bandgap and the localization of the topological edge state. The described behavior is clearly manifested in the simulations and experiment, since the localization of the edge state existing at nearly the same frequency indeed becomes less pronounced (see Figure~\ref{fig:fig3}b).

Finally, if all meta-atoms are aligned in the same way ($\alpha=0^\circ$), the bandgap closes up and no localized states are expected. This is exactly the case found in the simulated and measured data in Figure~\ref{fig:fig3}c, where we do not observe any localization in the entire frequency range of interest.

\begin{figure}[b!]
\centering
\includegraphics[width=1.0\textwidth]{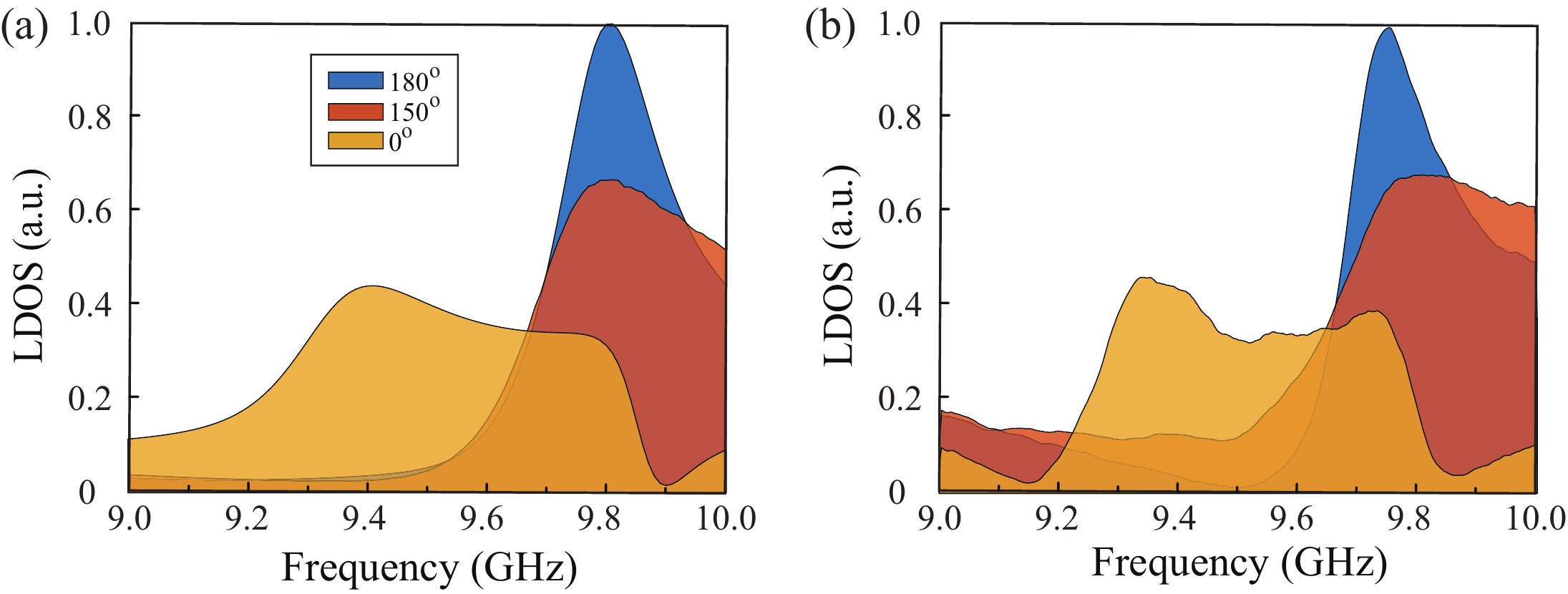}
\caption{{\bf Normalized local density of states (LDOS).} The results are extracted from (a) simulated and (b) experimentally measured data for the designed array with three different orientations of the meta-atoms. The plotted quantity is proportional to $|E_y|^2$ above the central element of the array.}
\label{fig:fig4}
\end{figure}

To identify the frequency of the topological edge state more precisely, we post-process the data for the three array configurations retrieving the normalized local density of states (LDOS) at the interface of the array as a function of the excitation frequency. The LDOS is calculated as a ratio of the squared amplitude of the $y$-component of the electric field $|E_y|^2$ summed over all points in the area located above the single central element to the total value of this field summed over all elements in the array. Then the obtained LDOS is normalized to the maximal value of $|E_y|^2$ extracted from the acquired data set. The results retrieved from the simulated and experimental data are plotted in Figures~\ref{fig:fig4}a and \ref{fig:fig4}b, respectively. Pronounced LDOS peaks for $\alpha=180^\circ$ and $\alpha=150^\circ$ hint towards the formation of the localized interface state with the frequency $9.75$~GHz. At the same time, LDOS for $\alpha=0^\circ$ does not feature any pronounced maximum indicating the absence of the interface state.

\begin{figure*}[t!]
\centering
\includegraphics[width=1\textwidth]{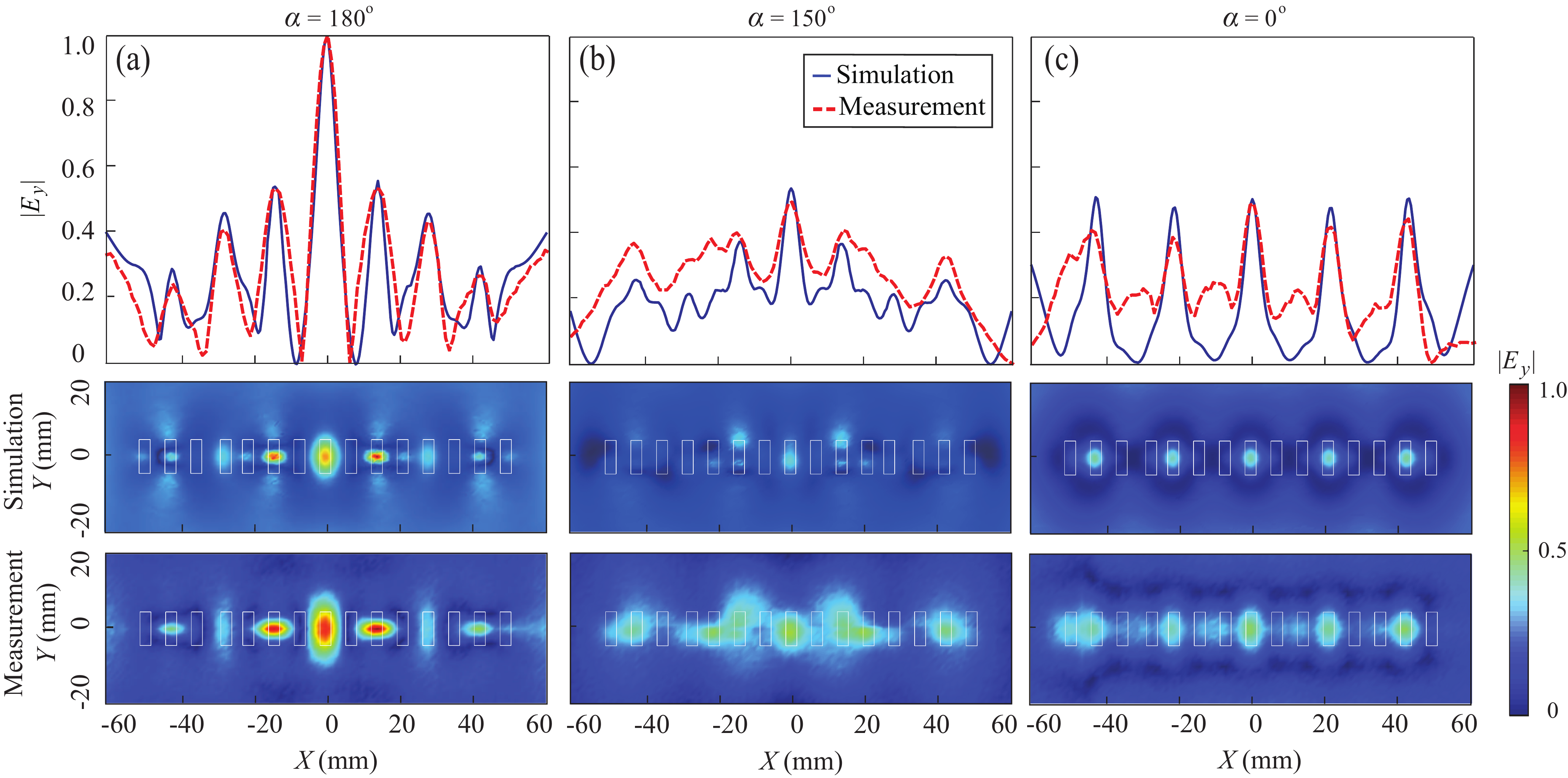}
\caption{{\bf Simulated and measured characteristics of the electric near-field at the specified frequency of the interface state.} Magnitude profiles (top row), simulated (middle row) and measured (bottom row) patterns of the $y$-component of the electric near-field ($|E_y|$) at the fixed frequency of 9.75~GHz for three array configurations that differ by several meta-atoms rotated on the angle (a) $\alpha=180^\circ$, (b) $150^\circ$, and (c) $0^\circ$. The mapping is performed at a distance of 1.5 mm above the structure plane. Parameters of the array are indicated in the caption to Figure~\ref{fig:fig1}.
}
\label{fig:fig5}
\end{figure*}

Focusing on the estimated frequency of the topological state, we analyze the patterns of the electric near-field distribution presented in Figure~\ref{fig:fig5}. In line with the previous discussion, we observe that the interface state localization gradually deteriorates when we decrease the rotation angle $\alpha$ from $180^\circ$ to $150^\circ$ and lower. Once $\alpha$ reaches $0^\circ$, the topological state fully disappears. Note that the results of numerical simulations fit the experimental ones quite closely which confirms high quality of the fabricated sample. 

\section{Conclusions}

In conclusion, we have experimentally demonstrated a strategy to continuously tune the localization of the topological interface state in a one-dimensional array with equidistant spacing. In our experiments, the mid-gap interface state trapped by a topological defect was excited by plane-wave-like radiation from the far field and imaged in the near field, revealing its resonant dipolar nature and characteristic staggered profile. Our idea to exploit the rotation degree of freedom of non-centrosymmetric meta-atoms can also be extended to the lattices of higher dimensionality paving a way towards higher-order topological systems and suggesting novel opportunities for the design of reconfigurable all-dielectric metadevices.

\section{Acknowledgements}

Theoretical models and numerical simulations were supported by the Russian Science Foundation (Grant No.~21-79-10209). D.A.S. acknowledges support from the Australian Research Council (DE190100430). V.R.T. acknowledges Jilin University’s hospitality and financial support.

\bibliography{ms}

\end{document}

\title{Reconfigurable topological states in arrays of bianisotropic particles \\ {\small Supporting Information}}

\author{Zuxian~He$^*$}
\affiliation{State Key Laboratory of Integrated Optoelectronics, College of Electronic Science and Engineering, International Center of Future Science, Jilin University, 2699 Qianjin Street, Changchun 130012, China} 
\thanks{Authors who contributed equally to this work}

\author{Daniel A. Bobylev$^*$}
\affiliation{School of Physics and Engineering, ITMO University, Saint Petersburg, Russia}
\thanks{Authors who contributed equally to this work}

\author{Daria A. Smirnova$^*$}
\affiliation{Research School of Physics, Australian National University, Canberra ACT 2601, Australia}
\affiliation{Institute of Applied Physics, Russian Academy of Science, Nizhny Novgorod, Russia}
\thanks{Authors who contributed equally to this work}

\author{Dmitry~V.~Zhirihin}
\affiliation{School of Physics and Engineering, ITMO University, Saint Petersburg, Russia}

\author{Maxim A. Gorlach}
\affiliation{School of Physics and Engineering, ITMO University, Saint Petersburg, Russia}
\email{m.gorlach@metalab.ifmo.ru}

\author{Vladimir R. Tuz}
\affiliation{State Key Laboratory of Integrated Optoelectronics, College of Electronic Science and Engineering, International Center of Future Science, Jilin University, 2699 Qianjin Street, Changchun 130012, China}

\maketitle

\section{Dipole modes of a single bianisotropic meta-atom}\label{sec:Single}

To get a qualitative picture of hybrid dipole mode formation, we first analyze the modes of a single high-index ceramic disk (Figure~S1a). Geometric and material parameters of the disk correspond to the commercially available high-permittivity microwave ceramics: $\varepsilon = 24.8$, $\tan \delta \approx 10^{-3}$, $r = 4.5$ mm, and $h = 3$ mm. The results of full-wave numerical simulation of eigenmodes are provided in Figure~S1b. 

\begin{figure}
\centering
\includegraphics[width=0.8\linewidth]{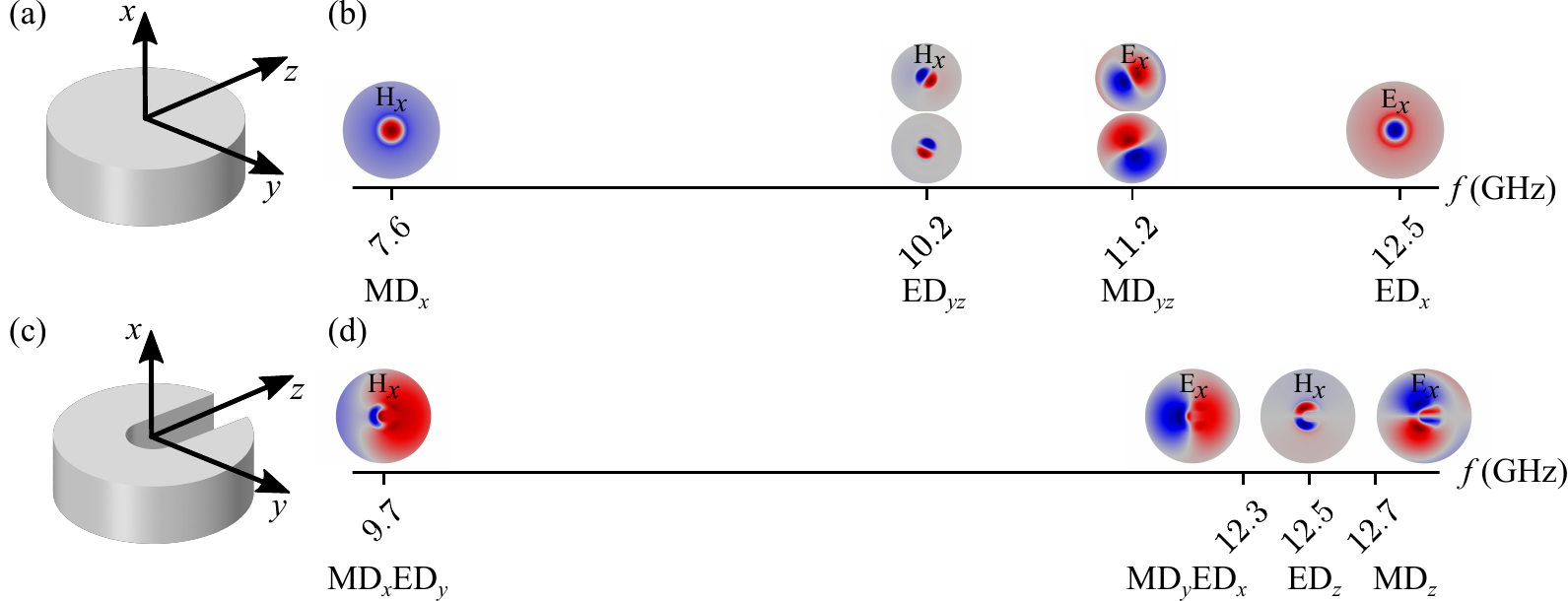}
\caption{{\bf Design of bianisotropic meta-atom.} (a) The initial non-bianisotropic (unperturbed) particle. (b) Dipole eigenmodes of a single unperturbed meta-atom (a). Each mode is represented either by pure electric or pure magnetic dipole. (c) Bianisotropic (perturbed) meta-atom. (d) Dipole eigenmodes of a single perturbed meta-atom. The lowest-frequency mode contains both $y$-oriented electric and $x$-oriented magnetic dipole moments, which have comparable amplitudes ($|p_y|/|m_x| \approx 0.86$).}
\label{fig:S1}
\end{figure}

Due to the preserved spatial inversion symmetry of the disk, its dipole modes contain either pure electric, or pure magnetic multipole moment of order $l=1$. If spatial inversion symmetry is broken, the electric and magnetic multipole moments of order $l=1$ mix giving rise to bianisotropy, or magneto-electric coupling. As a result, the response to the external electromagnetic fields $\mathbf{E}$, $\mathbf{H}$ is captured by the equations
\begin{equation}
\begin{gathered}
\mathbf{p} = \hat{\alpha}^{(\mathrm{ee})} \mathbf{E} + \hat{\alpha}^{(\mathrm{em})} \mathbf{H}, \\
\mathbf{m} = \hat{\alpha}^{(\mathrm{me})} \mathbf{E} + \hat{\alpha}^{(\mathrm{mm})} \mathbf{H}.
\end{gathered}
\end{equation}
Here, $\mathbf{p}$/$\mathbf{m}$ are the electric/magnetic dipole moments, $\hat{\alpha}^{(\mathrm{ee/mm})}$ are the conventional electric/magnetic polarizability tensors and $\hat{\alpha}^{(\mathrm{em/me})}$ are the magneto-electric coupling tensors. The explicit structures of polarizability and magneto-electric coupling tensors are defined by the type of the meta-atom symmetry breaking. In this work, we consider the mirror symmetry breaking in $xy$ plane (Figure~S1c), which ensures sufficient spectral separation of the hybrid dipole mode from the rest of the modes (Figure~S1d) and allows to construct a dynamically reconfigurable array (see Figure~1a of the manuscript), where the disks can be easily rotated. A simple symmetry analysis together with time-reversal symmetry leads to the following structure of the tensors:
\begin{equation}
\hat{\alpha}^{(\mathrm{ee/mm})} =
\begin{pmatrix}
\alpha^{\mathrm{e/m}}_1 & 0 & 0 \\
0 & \alpha^{\mathrm{e/m}}_2 & 0 \\
0 & 0 & \alpha^{\mathrm{e/m}}_3
\end{pmatrix}, \ 
\hat{\alpha}^{(\mathrm{em})} = (\hat{\alpha}^{(\mathrm{me})})^{\dagger} = 
\begin{pmatrix}
0 & i\kappa_1 & 0 \\
-i\kappa_2 & 0 & 0 \\
0 & 0 & 0
\end{pmatrix}.
\end{equation}
Thus, the particle supports hybrid dipole modes containing both $y$-oriented electric and $x$-oriented magnetic dipole moments.

Next we focus on a specific experimentally relevant design (Figure~S1c) and optimize its geometric parameters numerically to ensure substantial mixing of the electric and magnetic dipole moments ($|p_y|/|m_x| \approx 0.86$) in the lowest-frequency hybrid dipole mode. Such hybridization leads to strong dependence of the interaction between these particles on their mutual orientation as discussed in the Section~II below.

\section{Effective coupling of the meta-atoms and design of topological array}

\begin{figure}[t]
\centering
\includegraphics[width=0.8\linewidth]{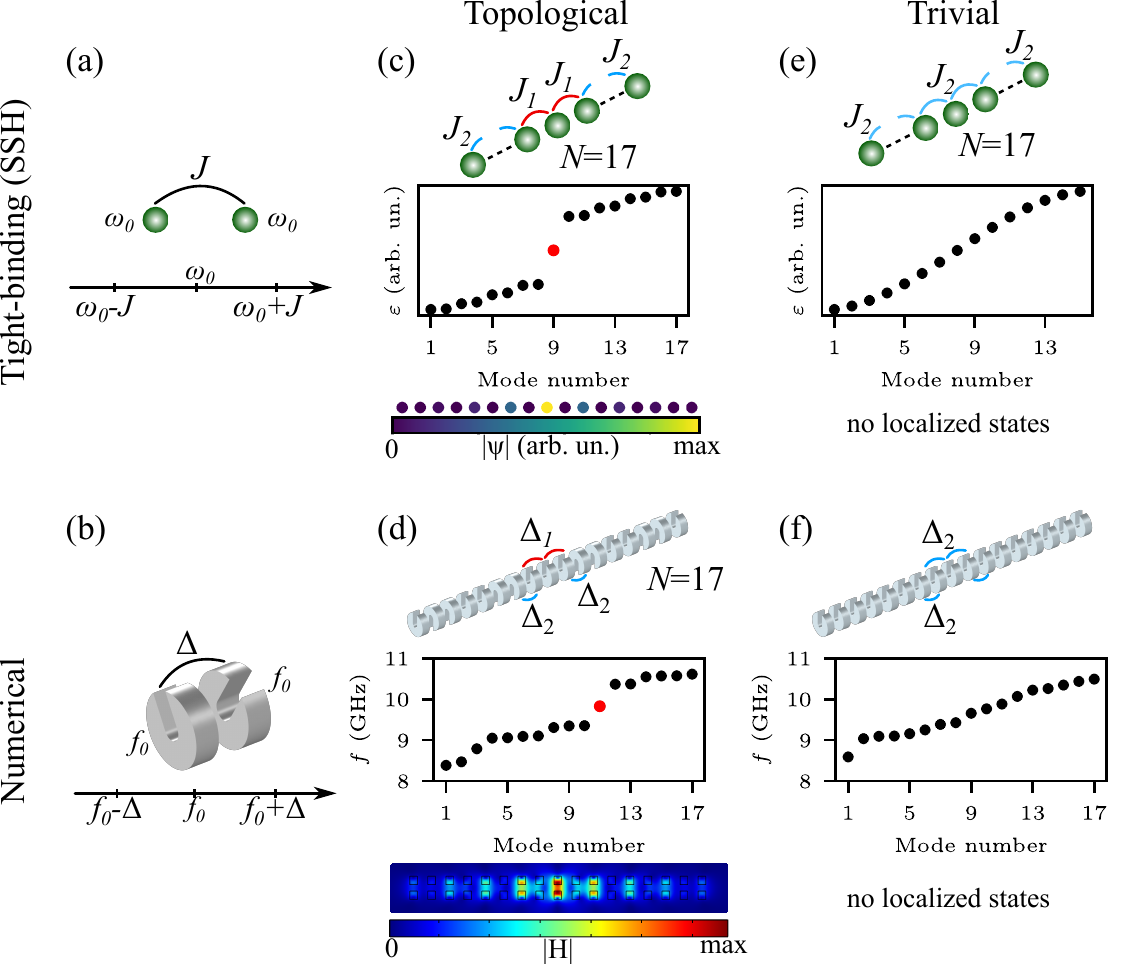}
\caption{{\bf Tailoring the coupling between bianisotropic meta-atoms.} (a) Splitting of the modes for the two sites with eigenfrequency $\omega_0$ coupled with the strength $J$. (b) Splitting of the lowest-frequency hybrid dipole mode for a pair of interacting bianisotropic meta-atoms. (c) Energies of the modes in a Su-Schrieffer-Heeger tight binding lattice with two repeated weak coupling links $J_1<J_2$ in the middle and total number $N=17$ of sites. A localized interface state is observed.  (d) Su-Schrieffer-Heeger-type array of bianisotropic meta-atoms with the orientation-dependent coupling. An in-gap interface-localized mode is observed. (e) Energies of the modes in a finite lattice with uniform coupling. No bandgap and localized states are formed. (f) An array of identical equidistant bianisotropic meta-atoms. Due to the uniform effective coupling, no localized modes appear.}
\label{fig:S2}
\end{figure}

As elaborated in the manuscript main text (see e.g. Figure~2c), the interaction of the meta-atoms is strongly sensitive to their mutual orientation at fixed distance. Hence, if the meta-atoms comprising the array have different orientations, their effective couplings will vary even if the lattice itself is equidistant. At the same time, as demonstrated in Section~\ref{sec:Single}, the lowest-frequency hybrid dipole mode of a single meta-atom is spectrally separated from the higher-frequency modes by a gap around $2.6$~GHz.

Given these circumstances, the array of such meta-atoms can be described by the effective tight-binding model, where sites correspond to the lowest-frequency hybrid mode of the respective meta-atoms, while the effective couplings are recovered from the matrix elements of the dyadic Green's function responsible for their interaction. This is schematically shown in Figure~\ref{fig:S2}a: the initial mode with frequency $\omega_0$ splits into two new modes with frequencies $\omega_0-J$ and $\omega_0+J$ due to interaction.

As a first approximation, we incorporate in our simplified model only the coupling of the neighboring particles ignoring the interaction between more distant disks. As a result, we immediately recover the well-celebrated Su-Schrieffer-Heeger model~\cite{Su} which consists of the sites with the same eigenfrequency and alternating nearest-neighbor couplings (Figures~\ref{fig:S2}c and \ref{fig:S2}e). This model is known to possess nontrivial Zak phase~\cite{Zak} that gives rise to topological edge states and interface states. In our tight-binding calculations in Fig.~\ref{fig:S2}, the chosen number of sites ensures the absence of topological edge states and only interface states appear.

In reality, the designed system features more complicated physics including long-range couplings of the meta-atoms, their higher-frequency modes and radiative losses. However, an important feature of topological invariants is that they cannot change continuously, but only abruptly at the topological transition point. Therefore, if the two models (simplified and the real one) have topologically equivalent band structures, they possess the same topological invariants. Based on this analysis, we expect the topological states to appear in our electromagnetic system as well, which is confirmed by the full-wave numerical eigenmodes simulations (Figures~\ref{fig:S2}d and \ref{fig:S2}f). As expected, a gapped topological interface mode emerges at the weak-weak defect ($\Delta_1<\Delta_2$, Figures~\ref{fig:S2}d).

\bibliography{supplement}